\documentclass[conference,compsoc]{IEEEtran}
\IEEEoverridecommandlockouts
\usepackage{cite}
\usepackage{amsmath,amssymb,amsfonts}
\usepackage[normalem]{ulem}
\usepackage{algorithmic}
\usepackage{graphicx}
\usepackage{textcomp}
\usepackage{xcolor}
\usepackage[inkscapelatex=false]{svg}
\usepackage{listings}

\usepackage{tabularx}
\usepackage{booktabs}
\usepackage[font=footnotesize]{caption}
\usepackage{multicol}
\usepackage{xltabular}
\usepackage{cuted}
\usepackage{soul}
\usepackage{xtab}
\usepackage{afterpage}
\usepackage{ltxtable}

\usepackage{listings}
\usepackage{xcolor}

\def\BibTeX{{\rm B\kern-.05em{\sc i\kern-.025em b}\kern-.08em
    T\kern-.1667em\lower.7ex\hbox{E}\kern-.125emX}}
\begin{document}
\newcommand{\jelena}[1]{{\color{red}[Jelena: #1]}}
\newcommand{\sam}[1]{{\color{blue}[Sam: #1]}}
\newcommand{\brian}[1]{{\color{purple}[Brian: #1]}}
\newcommand{\old}[1]{{\color{gray}[Old/Pending changes: #1]}}

\title{CISAF: A Framework for Risk Assessment of Academic and Research Cyberinfrastructure}
% \title{CISAF: An Implementation Guide of Risk Management Framework on Academic and Research Cyberinfrastructure}
% {\footnotesize \textsuperscript{*}Note: Sub-titles are not captured for https://ieeexplore.ieee.org  and

% NEED TO ANONYMIZE BEFORE SUBMISSION
\author{\IEEEauthorblockN{Qishen Sam Liang}
\IEEEauthorblockA{\textit{Information Sciences Institute} \\
\textit{University of Southern California}\\
Los Angeles, United States \\
qishenl@alumni.usc.edu}
\and
\IEEEauthorblockN{Jelena Mirkovic}
\IEEEauthorblockA{\textit{Information Sciences Institute} \\
\textit{University of Southern California}\\
Los Angeles, United States \\
mirkovic@isi.edu}
\and
\IEEEauthorblockN{Brian Kocoloski}
\IEEEauthorblockA{\textit{Information Sciences Institute} \\
\textit{University of Southern California}\\
Los Angeles, United States \\
briankoco@gmail.com}

}

\maketitle

\begin{abstract}
Academic and research cyberinfrastructures (AR-CIs) present unique risk and security challenges due to their collaborative nature and heterogeneous components. However, the industry-well-accepted risk assessment approaches have limited compatibility and specificity for complex systems like cyberinfrastructures (CIs). We propose CyberInfrastructure Security Analysis Framework (CISAF) -- a simple, systematic, mission-centric, and integrable approach to file for CI risk assessment. CISAF serves as a plug-in to traditional risk assessment/management frameworks, a convenient tool to extrapolate the cybersecurity risks for complex cybersystems. It guides AR-CI administrators through a top-down process: (1) defining unacceptable losses, (2) identifying associated system hazards and critical assets, (3) analyzing possible attack paths targeting these critical assets, (4) cross-referencing security mechanisms deployed along these attack paths, and (5) integrating results into established risk assessment/management frameworks. By combining information about CI architecture, missions, attack graphs, and security mechanisms, CISAF provides actionable guidelines, structured views of high-priority security risks, and streamlined translation into established frameworks.
\end{abstract}


\section{Introduction}

\textbf{A}cademic and \textbf{r}esearch \textbf{c}yber\textbf{i}nfrastructures (AR-CIs, for short) are complex networks, with integrated hardware, software, services, and datasets, which are developed and used by different research communities to advance science and facilitate open access to scientific resources~\cite{nsf_cyberinfrastructure_2024}. Like any networked resource, AR-CIs are subject to outsider and insider attacks. But unlike many commercial networked systems, AR-CIs face specific challenges that come from sporadic funding and the open nature of their user base.

AR-CI funding is often small and time-limited, which limits the size and expertise of their DevOps staff and the effort they can invest in security. In addition, AR-CIs are highly motivated to support various use cases, some of which may create additional security risks. Further, AR-CIs may include nonstandard, Internet-enabled devices, such as Internet of Things (IoT) devices and automated lab equipment \cite{adams2022findings}. The complexity and security risks of an AR-CI increase with increased device diversity. Lastly, AR-CIs often run custom management software, which may not be sufficiently secure~\cite{adams2021state}.

There is a need for practical and easy-to-follow guidelines for AR-CIs to create a risk management profile. Among the existing cybersecurity frameworks for IT and OT systems, NIST’s standard is one of the widely accepted security guidelines. NIST SP 800-30~\cite{jtf2012guide} and NIST SP 800-160 \cite{ross2021developing} are two cybersecurity frameworks for Internet-enabled systems and infrastructures. Both documents provide high-level analytic practices and processes that are applicable to IT and OT systems, but they are not immediately and directly applicable to AR-CIs. The ATT\&CK \cite{mitre_attack_2025} and STRIDE \cite{howard2006security} frameworks provide structured approaches to enumerate threats for single devices or services, but they cannot be readily extended to an AR-CI's scale and complexity. Mission Based Cyber Risk Assessments (MBCRAs) \cite{denaray2022analysis}, System-Theoretic Process Analysis for Security (STPA-Sec) \cite{young2021intro, sahay2023comparative}, and MISSION AWARE \cite{bakirtzis2017mission} provide mission-based security analysis for Cyber-Physical Systems (CPS), though insightful, they are not directly transferable to be applied on AR-CIs. The Cyber Analysis Visualization Environment (CAVE) \cite{pecharich2016mission} provides a security framework for mission-critical systems, but it is too complicated to implement and does not necessarily fits many AR-CIs. All the security frameworks gathered have limitations when analyzing AR-CIs.

We propose the Cyber-Infrastructure Security Analysis Framework (CISAF)--a simple, systematic, mission-centric, and integrable approach to structurally analyze the security posture of a CI as a plug-in for the commonly accepted guideline. CISAF makes the following contributions:

\begin{itemize}
    \item \textbf{Customized for Each AR-CI.} AR-CIs are designed and constructed by different teams and organizations; they serve different purposes, and thus have different security challenges. CISAF starts with the high-level architecture of an AR-CI and its mission, and customizes the rest of the analysis to these elements.

    \item \textbf{Easy to use and actionable.} The CISAF assessment is easy for AR-CI teams to understand and apply. It contains step-by-step analysis and produces transferable results, which ultimately lead to a risk management profile and actionable mitigations.

    \item \textbf{Integrable to Popular Frameworks} CISAF serves as a component and a detailed instruction for the widely accepted security framework, NIST SP 800-30~\cite{jtf2012guide}, as demonstrated in this paper.

     \item \textbf{Allows for Prioritization.} CISAF analysis provides critical information for analysts to reason about different cybersecurity improvements, and select those that make the biggest reductions in their cyber risk. 
\end{itemize}

CISAF complements existing analysis frameworks. Structurally it is inspired by and adopts elements of MISSION CENTRIC~\cite{bakirtzis2017mission}, MBCRA~\cite{denaray2022analysis}, and CAVE~\cite{pecharich2016mission}, specifically a mission-centric approach for prioritization and attack graph construction. CISAF should be supported with a security analysis approach, such as ATT\&CK~\cite{mitre_attack_2025} and STRIDE~\cite{howard2006security}, to construct the possible attack strategies. 

\section{Related Work}

Trusted CI proposed a cybersecurity program implementation guideline for researchers, which contains four main ``pillars''--Mission Alignment, Governance, Resources, and Controls--and sixteen ``Musts'' for Research CI Operators \cite{jackson2021trusted_ci}. The insights are valuable, but the guideline is high-level, and focus more on teaching CI researchers how to reason about cybersecurity than on analyzing the security posture. Further, the guideline does not specify a step-by-step, actionable security and threat analysis methodology, although it strongly upholds the understanding of assets through documentation and classification.

Aside from their program implementation guidelines, Trusted CI also published reports that analyze the security posture of various CIs \cite{adams2021state, adams2022findings}. In the CI software project report \cite{adams2021state}, Trusted CI interviewed CI project teams and highlighted several bad practices for CI software projects, such as the lack of threat modeling, documentation, maintenance, vulnerability tracking, prioritization, and poor coding language choices. Similarly, Trusted CI's Operational Technology report \cite{adams2022findings} highlights the OT cybersecurity shortcomings throughout the process of digitizing and making OT technology smart. 

NIST has provided guidance that shaped cybersecurity practices in various IT and OT system designs. The widely recognized Cybersecurity Framework (CSF) 2.0 \cite{nist_csf2_2024} offers a high-level structure organized around the six core functions (Govern, Identify, Protect, Detect, Respond, and Recover) that helps organizations manage and communicate cybersecurity risk across their lifecycles. Foundational to the CSF 2.0 ``Identify'' is the cybersecurity risk assessment, which refers to the Guide for Conducting Risk Assessments (NIST SP 800-30), analyzing threats, vulnerabilities, and impacts \cite{jtf2012guide}. 

Applying these foundational NIST standards presents practical challenges for teams assessing the security posture within AR-CIs: The high-level, organization-wide perspective of the CSF 2.0 often requires significant interpretation, knowledge, research, and time, and is often too big to start. 

Many security frameworks provide procedural guidelines for security analysis of specific types of cybersystems, at a fine granularity. The STPA-Sec~\cite{young2021intro,sahay2023comparative} and MISSION AWARE~\cite{bakirtzis2017mission} were originally designed for cyber-physical systems (CPS). These frameworks are not well-suited to IT-heavy CIs. Further, they are designed to be applied during the system design phase, following the principle of ``secure by design'' rather than providing continuous security estimation and support. Even though some aspects could be applied during operations, they are very comprehensive and would take significant time and human resources \cite{denaray2022analysis}. CAVE~\cite{pecharich2016mission} is a very comprehensive and fine-grained analysis framework for critical cybersystems, such as the system for space missions. CAVE is highly integrated and automated, pulling in data about network architecture, installed software, and security mechanisms from different existing databases. Such analysis is prohibitively costly for an AR-CI to conduct. 

Aside from the security frameworks designed for cybersystems, there are numerous security models and knowledge bases, such as MITRE ATT\&CK~\cite{mitre_attack_2025} and STRIDE~\cite{howard2006security}, for per-device security analysis. They can complement the CISAF approach to enumerate attacks and find vulnerabilities in AR-CIs.

\section{Background}

AR-CIs are complex and interconnected ecosystems comprising a variety of resources (software, hardware, network elements, data). They provide different levels of access and privileges to various research communities and individuals, thereby enabling significant scientific discoveries across various domains~\cite{nsf_cyberinfrastructure_2024}.

The architecture of AR-CIs varies considerably regarding user entry points, the mission of the CI, and the internal distribution of critical assets and sensitive data. Correspondingly, the degree of control granted to users differs across CIs, spanning from limited actions such as read-only file retrieval to privileged operations like superuser access or direct interaction with physical systems. For example, the SPHERE testbed \cite{sphere_colab} provides Virtual Machines (VMs) and bare-metal machines for registered users; FABRIC \cite{fabric_knowledge_base_2023} and ACCESS \cite{access_allocations_2025} provide VMs for registered users; and SAGE \cite{sage_edge_apps_2025} allows uploading, building, and sharing of AI applications for deployment on Docker containers across the network of software-defined sensors with given APIs and schedulers. Further, ZEUS \cite{zeus_proposal_2025} runs a yearly application cycle, in which only the researchers with accepted proposals get to access the research CI, while projects like ngEHT \cite{ngeht_opportunities_2025} and Neutron Spin Echo for the Nation \cite{cns_neutron_2025} only provide access to limited researchers, having neither a public registration portal nor a publicly accessible dataset. Greater access privileges can facilitate more versatile research, but they increase the risk to a CI from cyber threats. 

AR-CIs contain several unique characteristics that distinguish their security landscape from enterprise IT and OT environments. One is the difference in mission and objectives. Commercial cybersystems are operated for profit and thus have more funding, more security measures, and more detailed monitoring of their critical assets. AR-CIs are often supported by small research funds, aiming to democratize access to resources, facilitate fast research processes and/or broaden educational benefits. Often, there is a smaller technical team, and thus less stringent control and monitoring of critical resources, potentially allowing attacks and exploitations to happen more easily and without detection. AR-CI administrators may also lack the knowledge and resources to develop and maintain comprehensive security systems and sensors to track usage and access patterns.

Another defining characteristic of AR-CIs is their strong focus on collaboration and openness inherent in research and education fields. Many AR-CI missions involve open sharing of research platforms, research outputs, and resources across multiple institutions or nations. The necessity of sharing data, computational resources, and research processes among different organizations and teams creates tensions with traditional security principles that prioritize restriction and control. 

Lastly, research activities on an AR-CI often develop new ways of leveraging resources or data, introducing additional security challenges. These research activities may themselves use outdated or custom software and services, with limited support and possibly unknown vulnerabilities, requiring a highly adaptable and updated security framework to adjust to emerging security threats.

\section{CISAF Approach}

\begin{figure}[htbp]
  \centering
  \caption{The workflow of CISAF}
  \includegraphics[width=0.55\linewidth]{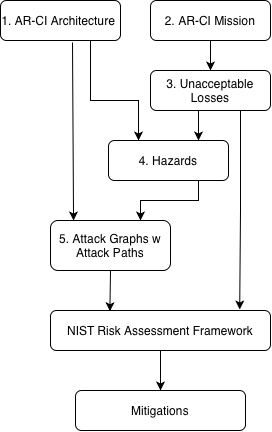}

  \label{fig:flow}
\end{figure}

The \textbf{C}yber \textbf{I}nfrastructure \textbf{S}ecurity \textbf{A}nalysis \textbf{F}ramework or CISAF, a systematic, mission-centric, and integrable framework, aims to represent and reason about the security posture of AR-CIs. CISAF starts with a \textit{simplified functional architecture} of the given AR-CI, supplied by the CI team, and a definition of the AR-CI's \textit{mission}. It proceeds to define adverse events that would jeopardize the AR-CI mission, called \textit{unacceptable losses}. Next, it defines \textit{hazards}, attacker-driven events on the AR-CI architecture that may cause one or multiple losses. Next, it develops \textit{attack graphs} consisting of \textit{attack paths}, each describing a series of hops the adversary has to take to create a hazard. Next, CISAF analyzes the attack techniques and protections along the attack paths--this information helps AR-CI teams reason about how these mechanisms disable or weaken the attack paths. Those attack paths that have few or no security mechanisms are candidates for mitigation actions. Finally, CISAF concludes the results and populates the Adversarial Risk Table, helping researchers to complete the Risk Management framework. Figure~\ref{fig:flow} illustrates CISAF's workflow.

\subsection{Simplified AR-CI Architecture}

CISAF starts with a high-level, functional architecture of an AR-CI, annotated with critical assets. The architecture includes important servers and services in the AR-CI, their physical and logical connections, physical elements that are accessible by AR-CI users or outsiders, and critical data assets, such as user credentials and experimental data. Physical or logical components that have identical functionality are grouped together in the architecture.

\begin{figure*}[htbp]
  \centering
    \caption{SPHERE's architecture, annotated with user access paths and data/service assets. Blue, dashed lines show possible user data flows. }
  
  \includegraphics[width=0.8\linewidth]{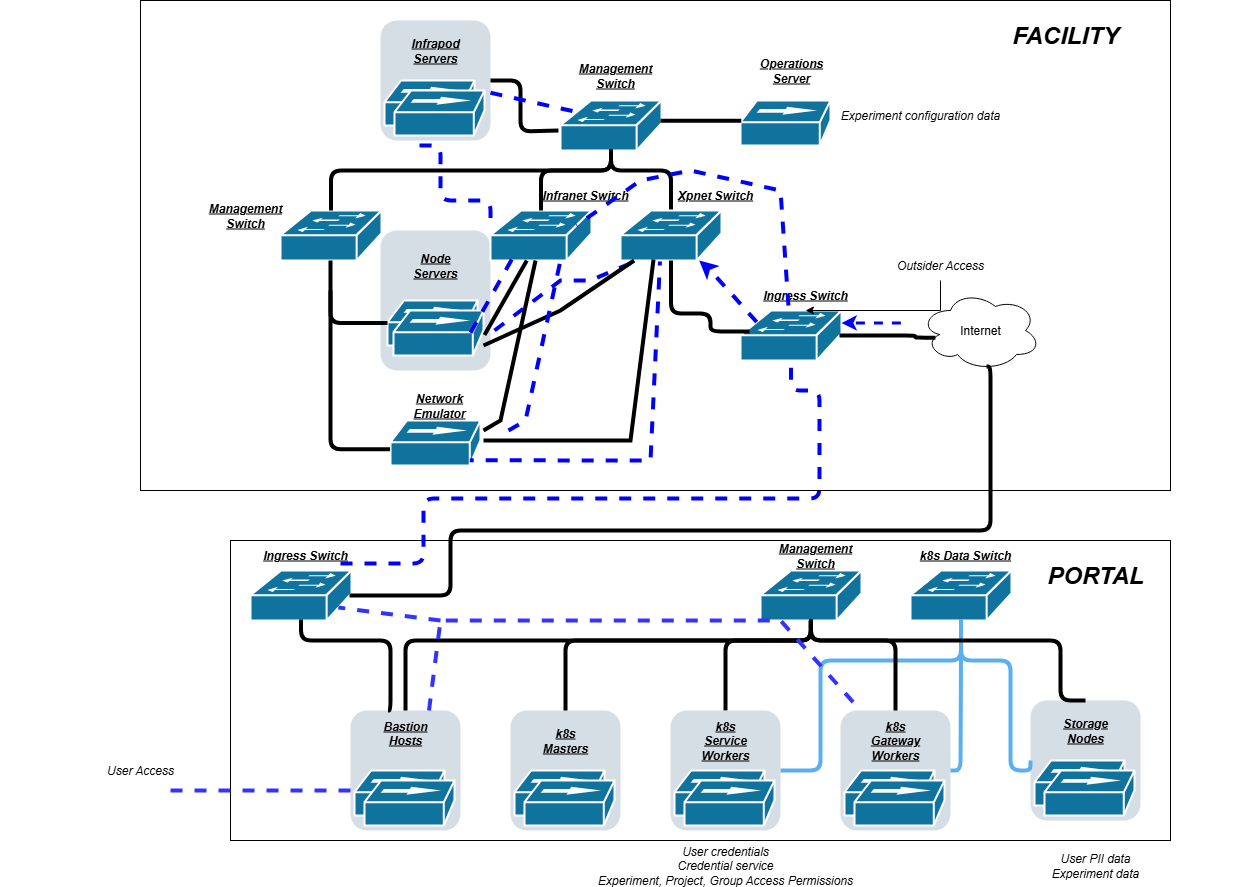}

  \label{fig:spherekg}
  % \Description{A partial view of the SPHERE testbed's infrastructure, detailing the ways the experiment-related nodes are connected.}
  % \jelena{where are XDCs in this figure}

\end{figure*}

\textbf{SPHERE Testbed Architecture.} We illustrate this step on the SPHERE Testbed. Figure~\ref{fig:spherekg}
shows SPHERE's architecture, annotated with data and service assets. There are two parts of the infrastructure: a \textit{portal}, which manages user PII data and credentials, group, experiment, and project access rights, and credentials that allow user access to experiments; and a \textit{facility}, which allocates resources for experiments, including virtual and bare-metal machines, and network resources. Experiment nodes can freely reach the Internet via an infrapod---a rootless \texttt{Podman} pod dynamically orchestrated by \texttt{cniman} on the Fedora CoreOS Infrapod Server---and the ingress switch. Users can also configure their experiments to support external ingress via this dedicated channel. Users access experiments via the bastion server and via a user-specific, virtual gateway server.  This access between the gateway server and the experiment nodes occurs via a number of intermediate switches and the infrapod server, which creates a VLAN between a user's gateway server and their experiment nodes.

%\brian{Is that the goal of the KG? I thought the goal of the KG is simply to model the semantic relationships between things (e.g., whether a vulnerability affects an asset or not). What you've listed seem like goals of the whole system, not the KG}

\subsection{AR-CI Mission and Losses}

%\brian{I like this section}

A mission-centric approach helps resource-constrained organizations, such as AR-CI teams, to direct limited security resources exclusively toward critical assets \cite{pecharich2016mission,sahay2023comparative,bakirtzis2017mission}. 

In this step, the AR-CI teams should define their mission before entering threat analysis. 
Most AR-CIs aim to serve their research community, and can start with this statement. The statement should then be expanded, based on what kind of services and resources the AR-CI offers to users, and what access it facilitates.
 
 \textbf{SPHERE Testbed mission.} The SPHERE Testbed offers superuser access to virtual and bare-metal machines, as well as custom networks between nodes and Internet-facing networks. Consequently, we define SPHERE's mission in three parts: securing user data and experimentation; containing malicious outbound traffic; and protecting users' access to the testbed via web portal, CLI, and public internet port.

We derive \textit{unacceptable losses} directly from the mission definition. These are events that would endanger the CI's mission, by either jeopardizing its ability to serve its users, inflicting harm on users, or lowering the CI's reputation. A clear hierarchical order of losses could be listed, cross-referencing the core principles of system security (Confidentiality, Integrity, and Availability) in addition to the organization and project guidelines. An ordered listing of losses ensures that downstream hazard analysis and security controls target the most catastrophic operational failures first.

\textbf{SPHERE Testbed losses.} We identified five losses for SPHERE (Table~\ref{tab:losses}), ordered from highest to lowest priority: experimental data leak, experimental data loss, experimental data corruption, experiment interruption, and infrastructure misuse. Data leaks are ranked the highest, because their damage cannot be mitigated – once data is stolen, it is forever open to misuse by others. Experimental data loss occurs when the user loses access to experimental data, e.g., through deletion or ransomware. Data loss is concerning if it occurs long after the experiment has concluded. It is possible that users (usually graduate students) who conducted experiments and collected the data have left the project, and this makes recovery (e.g., via new experimentation) very costly. Similarly, experimental data corruption can have a large impact on a research project, especially if it is stealthy or hard to detect. Experiment interruption is less concerning, because a running experiment can always be restarted on the SPHERE Testbed, though it creates an unpleasant user experience. Finally, infrastructure misuse is of concern, because it could bring reputational damage and a waste of resources to the SPHERE Testbed. It is ranked the lowest, because it does not directly jeopardize SPHERE’s mission of supporting research, while other losses do.

% experimental data leak, experimental data loss, experimental data corruption, experiment interruption, and infrastructure misuse. Data leaks are ranked the highest, because their damage cannot be mitigated -- once data is stolen, it is forever open to misuse by others. Experimental data loss occurs when the user loses access to experimental data, e.g., through deletion or ransomware. Data loss is concerning if it occurs much after the experiment has concluded. It is possible that users (usually graduate students) who conducted experiments and collected the data have left the project, and this makes recovery (e.g., via new experimentation) very costly. Similarly, experimental data corruption can have a large impact on a research project, especially if it is stealthy and hard to detect, and if it occurs much after the experiment has concluded. Experiment interruption is less concerning, because a running experiment can always be restarted on SPHERE Testbed. Finally, infrastructure misuse is of concern, because it could bring reputational damage to SPHERE Testbed. It is ranked the lowest, because it does not directly jeopardize SPHERE's mission of supporting research, while other losses do.

\begin{table}[htbp]
  \caption{Unacceptable Losses}
  \begin{tabular}{ c|l }
    \toprule
    Loss Identifier & Description\\
    \midrule
    L1 & Experimental data leak\\
    L2 & Experimental data loss\\
    L3 & Experimental data corruption
    \\
    L4 & Interruption/termination of experiments \\
    & (node destroyed, node restarted)\\
    L5 & Misuse of infrastructure \\
    & (e.g., crypto mining, botnet for attacking others)\\
  \bottomrule
\end{tabular}
\label{tab:losses}
\end{table}

\subsection{From Losses to Hazards}

A ``hazard'' is a set of adversary actions on the CI that enables the adversary to create a loss. 
For example, a leak of experimental data (loss) can occur if an adversary gains unauthorized access to a file system holding the experimental data (hazard). 
Experiment Interruption (loss) can occur if an adversary conducts a Denial of Service (DoS) attack on the experiment network (Expnet) switch and disables the network during another experiment's run (hazard). The AR-CI team combines the \textit{unacceptable losses} with the knowledge captured from \textit{AR-CI architecture} to identify hazards of interest.

\textbf{SPHERE hazards.} Table \ref{tab:hazards} lists the hazards we identified on SPHERE for the losses in Table~\ref{tab:losses}.

% \afterpage{\clearpage
\begin{table*}[htbp]
% \onecolumn
\caption{A summary of hazards, their descriptions, and associated losses.} 
\label{tab:hazards}

\begin{tabular}{p{4cm} p{10.3cm} p{2.2cm} }
    \hline\textbf{Hazard} & \textbf{Hazard Description} & \textbf{Associated Losses}\\
    \hline
    H1-facility breach &  &\\
    H1.1-operation server subversion & The attacker can allocate or deallocate any resources, gain access to all experiments and delete or corrupt data on the experimental nodes. They can also leak that data to the outside.  & L1, L2, L3, L4, L5\\
    H1.2-network emulator subversion & The attacker becomes a person in the middle for all experimental communication. They can corrupt or drop network traffic. & L1, L2, L3 \\
    H1.3-infrapod server subversion & The attacker becomes a person in the middle for all communication between experiments and the outside, and can redirect, drop, or corrupt this traffic. They can also impersonate outside servers and gain the trust of users, leading to sensitive data loss.  & L1, L2, L3 \\
     H1.4-node server subversion & The attacker can change node images, interrupt experiments, and access experiments of other users that share the same physical node server. & L1, L2, L3, L4, L5\\
     H1.5-bastion server subversion & The attacker becomes a man-in-the-middle between the CI and the outside world. They can generate malicious traffic to the inside or to the outside, and interfere with experiments' outside connectivity, rendering the CI unusable. & L4, L5\\
    % \midrule
    \hline
    H2-portal breach &  & \\
    H2.1-storage server subversion & The attacker has access to long-term storage for all users and projects, and can delete or corrupt experimental data, or exfiltrate sensitive data. The attacker can interfere with project, group, and experiment membership to corrupt it, to delete it, or use it to gain additional privileges. & L1, L2, L3, L4, L5\\
   H2.2-identity server subversion & The attacker has access to all user information, such as credentials and PII. They can exfiltrate it or misuse it to impersonate a user and gain access to their resources. This access can be used to corrupt or delete experimental data, interrupt experiments, or misuse resources for unauthorized activities.& L1, L2, L3, L4, L5\\
    H2.3-certificate service subversion & The attacker can invalidate existing certificates and produce new ones, thus interfering with a user's access to experimental nodes, or giving the attacker unauthorized access to experiments. & L1, L2, L3, L4, L5\\
    % \midrule
    \hline
    H3-experiment breach & The attacker can interfere with the user's experiment, exfiltrate data, delete or corrupt experimental data, and misuse resources for profit or to attack others on the Internet. & L1, L2, L3, L4, L5\\
    % \midrule
    \hline
    H4-experiment impacted/impaired & Attacker impacted experimental or infranet network or devices, ie. DoSing experiment network, draining node server compute power, etc.. &L4 \\
    \hline
    H5-user turned malicious & An insider attacker starts misusing the infrastructure. & L5\\
    \hline

\end{tabular}

\end{table*}

\subsection{Attack Graph And Security Mechanisms}

Attack graphs investigate the steps an attacker must take---such as lateral movement, privilege escalation, and execution---to reach the target asset in an interconnected system \cite{10.1145/586110.586140}. The starting points of the graph are the nodes that the adversary has direct control over, such as the Internet, the portal k8s container, or an experiment node (VM or bare-metal). The ending points are the entities and assets that are involved in the hazards.

Attack graphs can be derived from architectural topology specifications. To lower subjective omissions during the analysis, CISAF recommends coupling architecture topology maps with a standardized threat knowledge base or framework. In this paper, we utilize the MITRE ATT\&CK Enterprise framework (STIX 2.1) \cite{mitre_attack_2025} on top of a multi-epoch agentic Large Language Model (LLM). We first create attack paths for specific critical assets, providing the documentation of the CI. The attack paths are then reviewed, appended to, adjusted, and fed into the second stage.

\label{sec:attack_graph_llm}
\textbf{LLM Threats Traversal.} Notice that the process can be done completely manually to fully avoid hallucinations, inconsistencies, and misjudgments. In fact, we advise against using LLMs alone for the analysis based on our experience, or it should at least be proctored, with researchers inspecting, correcting, and discussing with the model to achieve a higher correctness. In this paper, we use the LLM to establish a baseline. Detailed information about the LLM and the prompts could be found in section~\ref{sec:LLM} and Appendix~\ref{sec:appendix_prompt}.

From the LLM output, we filter for specific per-hop ATT\&CK tactics, such as \textit{Privilege Escalation}, \textit{Lateral Movement}, \textit{Initial Access}, and \textit{Impact}. 
% As these tactics track the adversary movement in the CI. The filtering applied to the last hop varies across hazards, and researchers may consider other tactics such as \textit{Credential Access}, \textit{Collection},\textit{Execution}, \textit{Exfiltration}, and \textit{Impact}. The result is then reviewed by one of the authors. 

\textbf{Human Auditing and Data Evaluation.} The result is reviewed and labeled by one of the authors. The current construct of our LLM exhibits a high false positive rate in auditing: Across the 3 attack graphs, 124 out of 226 parent techniques the LLM marked \textit{Exposed} (\texttt{[GAP]} or \texttt{[Partial]}) are labeled differently by the evaluator, while 5 out of 222 for \textit{Protected} (\texttt{[Covered]} and \texttt{[Omitted]}). Among the mislabeled \textit{Exposed} rows, the agent is likely to hypothesize non-existing high-risk ports on the user-facing network segment, potentially due to a small context window, and the design of the orchestration. Also, a design structure issue causes many rows to be mislabeled as \texttt{[GAP]} in the last hops for paths in Table~\ref{tab:paths3}. Since the model considers all techniques in multiple epochs, when one of the techniques gets system access to the destination node, most of the techniques from \textit{Impact} would be enabled by the root access and thereafter marked as \texttt{[GAP]} in the next epoch. The output produced by the LLM is a reasonable baseline for the researchers to step in and conduct further auditing. 

% Corresponding mitigation and detection strategies are provided along each ATT\&CK techniques~\cite{mitre_attack_2025}. 
% If some paths have none or insufficient security mechanisms to stop the traversal, the AR-CI team can focus their mitigation efforts on these paths. To prioritize mitigation, the AR-CI team can refer to the losses and attend first to those paths that lead to hazards associated with high-priority losses.

\textbf{SPHERE Testbed attack graphs and security mechanisms.} We will now illustrate the attack graphs for three hazards from Table~\ref{tab:hazards}: \uline{H1.3-infrapod server subversion} in Table~\ref{tab:hazard_h13_summary}, \uline{H1.1-operation server subversion} in Table~\ref{tab:hazard_h11_summary}, and \uline{H4-experiment impacted/impaired} in Table~\ref{tab:paths3} and relevant security mechanisms of SPHERE. All tables take the same construct, which simulates the adversary taking multiple possible entries and using tactics from the ATT\&CK framework, such as \textit{Initial Access}, \textit{Lateral Movement}, and \textit{Privilege Escalation}, depending on the hop to traverse to the critical assets. Hazard \uline{H4-experiment impacted/impaired} is slightly different as the last hop bifurcates into two situations. Depending on whether the adversary takes control of the Expnet/Infranet switch using \textit{Lateral Movement} before deploying the \textit{Impact} techniques.

\begin{table*}[t]
\centering
\footnotesize
\caption{Hazard H1.3 Threat Traversal Paths, Protections}
\label{tab:hazard_h13_summary}
\begin{tabularx}{\linewidth}{l X}
\toprule
\textbf{Path \& Origin} & \textbf{Protections}  \\
\midrule
\textbf{Path 1:} Experiment Node $\xrightarrow{\text{T1210}}$ Infrapod Pod $\xrightarrow{\text{T1068, T1611}}$ Infrapod Host Server & VLAN isolation, Unprivileged Podman containers, immutable Fedora CoreOS \\
\textbf{Path 2:} Experiment Node $\xrightarrow{\text{None}}$ Infranet Switch $\xrightarrow{\mathcal{S}_{\text{lateral}}}$ Infrapod Server & VLAN isolation, Out-of-band management plane, SSH auth \\
\textbf{Path 3:} External Network $\xrightarrow{\text{T1190}}$ Facility Ingress Switch $\xrightarrow{\mathcal{S}_{\text{lateral}}}$ Infrapod Server & Ingress filtering, Out-of-band management plane, SSH auth \\
\textbf{Path 4:} k8s container $\xrightarrow{\text{None}}$ Facility Switch / VXLAN $\xrightarrow{\mathcal{S}_{\text{lateral}}}$ Infrapod Server & K8s isolation, VXLAN isolation, Out-of-band management plane \\
\textbf{Path 5:} Experiment Node $\xrightarrow{\text{T1611}}$ Node Hypervisor  $\xrightarrow{\mathcal{S}_{\text{lateral}}}$ Infrapod Server & QEMU/KVM, Podman containers, SSH auth \\
\midrule
\multicolumn{2}{l}{\textbf{Technique Set Definitions}:} \\
% \multicolumn{2}{l}{\scriptsize $\bullet\ \mathcal{S}_{\text{escape}} = \{\text{T1037.004, T1053, T1055, T1068, T1078, T1098, T1543.005, T1546, T1548, T1611}\}$} \\
\multicolumn{2}{l}{\scriptsize $\bullet\ \mathcal{S}_{\text{lateral}} = \{\text{T1021, T1210, T1550, T1563, T1570}\}$} \\
% \multicolumn{2}{l}{\scriptsize $\bullet\ \mathcal{S}_{\text{lateral2}} = \{\text{T1021, T1072, T1080, T1210, T1534, T1550, T1563, T1570}\}$} \\
\bottomrule
% \multicolumn{2}{l}{\scriptsize \textit{See Appendix \ref{sec:appendix_h13} for full per-hop techniques and rationales.}}
\end{tabularx}
\end{table*}

\begin{table*}[t]
\centering
\footnotesize
\caption{Hazard H1.1 Threat Traversal Paths }
\label{tab:hazard_h11_summary}
\begin{tabularx}{\linewidth}{l X}
\toprule
\textbf{Path \& Traversal Chain} & \textbf{Protections} \\
\midrule
\textbf{Path 1:} Experiment Node $\xrightarrow{\text{T1611}}$ Node Hypervisor $\xrightarrow{\mathcal{S}_{\text{lateral}}}$ Ops Server & QEMU/KVM, Podman containers, SSH auth \\
\textbf{Path 2:} Bare-Metal Node $\xrightarrow{\text{None}}$ Access Switch $\xrightarrow{\mathcal{S}_{\text{lateral}}}$ Ops Server & VLAN isolation, Out-of-band management plane, SSH auth \\
\textbf{Path 3:} Experiment Node $\xrightarrow{\text{T1210}}$ Infrapod Pod $\xrightarrow{\text{T1068, T1611}}$ Infrapod Host & VLAN isolation, Podman containers, SSH auth\\
Server $\xrightarrow{\mathcal{S}_{\text{lateral}}}$ Ops Server & \\
\textbf{Path 4:} External Internet $\xrightarrow{\text{T1190}}$ Ingress Switch $\xrightarrow{\mathcal{S}_{\text{lateral}}}$ Ops Server & Ingress filtering, Out-of-band management plane, SSH auth \\
\textbf{Path 5:} Portal k8s container $\xrightarrow{\text{T1210, T1570}}$ Gateway Worker $\xrightarrow{\mathcal{S}_{\text{lateral}}}$ Ops Server & K8s isolation, VXLAN isolation, SSH auth \\
\midrule
\multicolumn{2}{l}{\textbf{Technique Set Definitions}:} \\
\multicolumn{2}{l}{\scriptsize $\bullet\ \mathcal{S}_{\text{lateral}} = \{\text{T1021, T1210, T1550, T1563, T1570}\}$} \\
% \multicolumn{2}{l}{\scriptsize $\bullet\ \mathcal{S}_{\text{external}} = \{\text{T1021, T1072, T1078, T1133, T1190, T1195, T1199, T1200, T1210, T1550, T1563, T1570, T1659}\}$} \\

\bottomrule
% \multicolumn{2}{l}{\scriptsize \textit{See Appendix \ref{sec:appendix_h11} for full per-hop techniques and rationales.}}
\end{tabularx}
\end{table*}

% The \textit{Protections} column in Table~\ref{tab:hazard_h11_summary} shows hypervisor boundaries and network segregation contains direct unauthorized intrusions. However, multi-hop traversals, though harder to achieve, reveals gaps ($S_{\text{lateral}}$) across shared management and control planes.

% \underline{Hazard H4-Experiment Impacted.} Finally, we look into attack paths for hazards H4-experiment impacted/impared. The attack graph is shown in Table~\ref{tab:paths3}.

\begin{table*}[t]
\centering
\footnotesize
% \label{tab:5}
\caption{Hazard H4 Threat Traversal Paths, Protections}
% \sam{[PENDING UPDATE, consider adding first Lateral then impact like Exp Node----\textit{lateral}---- Expnet----\textit{Impact}----Expnet]}}
\label{tab:paths3}
\begin{tabularx}{\linewidth}{l X}
\toprule
\textbf{Path \& Origin} & \textbf{Protections}  \\
\midrule
\textbf{$^1$Path 1A:} Experiment Node $\xrightarrow{\text{T1498, T1499}}$ Expnet Switch & VLAN \& VXLAN isolation, Out-of-band management plane \\
\textbf{$^2$Path 1B:} Experiment Node $\xrightarrow{\text{T1611}}$ Node Hypervisor $\xrightarrow{\text{T1498, T1499}}$ Infranet Switch & QEMU/KVM, Out-of-band management plane \\
\textbf{$^1$Path 2A:} Bare-Metal Node $\xrightarrow{\text{T1498, T1499}}$ Expnet Switch &  VLAN \& VXLAN isolation, Out-of-band management plane\\
\textbf{$^2$Path 2B:} Bare-Metal Node $\xrightarrow{\text{None}}$ Expnet Switch $\xrightarrow{\text{T1498, T1499}}$ Infranet Switch & VLAN \& VXLAN isolation, Out-of-band management plane \\
\textbf{$^2$Path 3:} Experiment Node $\xrightarrow{\text{T1210}}$ Infrapod Pod $\xrightarrow{\text{T1068, T1611}}$ Infrapod & VLAN isolation, Podman containers, \\
Host $\xrightarrow{\text{T1498, T1499}}$ Infranet Switch & Out-of-band management plane \\
\textbf{$^2$Path 4:} External Internet $\xrightarrow{\text{T1190}}$ Ingress Switch $\xrightarrow{\text{T1498, T1499}}$ Infranet/Expnet Switch & Ingress filtering, VLAN isolation, SSH auth \\
% \textbf{Path 4:} External Internet $\xrightarrow{\text{T1190}}$ Ingress Switch $\xrightarrow{\text{T1498, T1499}}$ Expnet Switch & Out-of-band management isolation, VXLAN VTEP encapsulation \\
\textbf{$^2$Path 5:} Portal k8s container $\xrightarrow{\text{T1210, T1570}}$ Gateway Worker $\xrightarrow{\text{T1498, T1499}}$ Infranet Switch & K8s isolation, VXLAN isolation, SSH auth \\
\midrule
\multicolumn{2}{l}{\textbf{$^{1,2}$Final hop Takeover Impact:}} \\
\textbf{Path Extra 1:} Non-Mangement Device $\xrightarrow{\text{None}}$ Expnet/Infranet & VLAN isolation, Out-of-band management plane\\
Switch $\xrightarrow{\mathcal{S}_{\text{impact}}}$ Expnet/Infranet Switch &  \\
\textbf{Path Extra 2:} Mangement Device $\xrightarrow{\text{T1021, T1210, T1550, T1563, T1570}}$ Expnet/ & Prior protections, SSH auth \\
Infranet Switch $\xrightarrow{\mathcal{S}_{\text{impact}}}$ Expnet/Infranet Switch &\\
% \multicolumn{2}{l}{\scriptsize $\bullet\ \mathcal{S}_{\text{impact}} = \{\text{T1485, T1486, T1489, T1490, T1491, T1495, T1496, T1498, T1499, T1529, T1531, T1561, T1565}\}$} \\
\midrule
\multicolumn{2}{l}{\textbf{Technique Set Definitions}:} \\
\multicolumn{2}{l}{\scriptsize $\bullet\ \mathcal{S}_{\text{impact}} = \{\text{T1485, T1486, T1489, T1490, T1495, T1496, T1498, T1499, T1529, T1531, T1561, T1565}\}$} \\

\bottomrule
% \multicolumn{2}{l}{\scriptsize \textit{See Appendix \ref{sec:appendix_h4} for full per-hop techniques and rationales.}}
\end{tabularx}
\end{table*}

\section{CISAF to Risk Assessment Standards}

While CISAF provides a domain-specific methodology for identifying attack paths and hazards within AR-CIs, these findings must often be translated into standardized frameworks for broader organizational risk management. 

Although CISAF can be used to populate multiple intermediate workbooks from NIST SP 800-30~\cite{jtf2012guide} (e.g., Table D-2 for Threat Sources, Table E-2 for Threat Events, and Table F-2 for Vulnerabilities), we map all findings directly to Table I-5. As per NIST SP 800-30 Appendix I, Table I-5 is the framework's master composite synthesis template---integrating Threat Sources (Cols 2--5), Threat Events (Col 1), Vulnerabilities (Col 8), Compounding Likelihood (Cols 7, 10, 11), and Impact (Col 12) into a single risk determination (Col 13). 

\begin{itemize}
    \item \textbf{Threat Event (Col 1):} Instantiated by the multi-hop attack paths on attack graphs.
    \item \textbf{Threat Source (Col 2):} the entry point's adversarial access level.
    \item \textbf{Threat Characteristics (Cols 3, 4, 5):} \textbf{Capability, Intent,} and \textbf{Targeting} are derived from the system model, technical privileges, tools, and objectives. For instance, authorized malicious users on Experiment Nodes are assumed to have \textit{High Capability} and \textit{Specific Targeting} toward infrastructure assets.
    \item \textbf{Relevance (Col 6):} \textit{Confirmed}, as analysis stems from \textit{Unacceptable Losses}.
    \item \textbf{Likelihood of Initiation (Col 7):} Evaluated via NIST SP 800-30 Table~G-2 based on adversary motivation, technical capability, and target asset attractiveness (\textit{e.g.}, access to privileged control planes or high-performance compute).
    \item \textbf{Vulnerabilities and Predisposing Conditions (Col 8):} Directly instantiated by CISAF's identified \texttt{[GAP]} and \texttt{[Partial]} technique evaluations (\textit{e.g.}, exposed SSH port) alongside predisposing architectural factors (\textit{e.g.}, tenant sudo access in Experiment Node).
    \item \textbf{Severity and Pervasiveness (Col 9):} Manually evaluated via NIST SP 800-30 Tables~F-2 and F-3 based on vulnerability exploitability and the difficulty of replicating the attack across the testbed.
    \item \textbf{Likelihood of Success (Col 10):}  Referencing NIST SP 800-30 Table~G-3, evaluate the length of the \textit{Attack Paths}, the number of \textit{Attack Techniques} deemed possible, and the \textit{Protections} (Security Mechanisms) listed. The success of an attack path requires success at every preceding boundary crossing. 
    Robust isolation mechanisms (\textit{e.g.}, QEMU/KVM virtualization and management network separation) enforce Low conditional success probabilities at critical hops.
    \item \textbf{Overall Likelihood (Col 11):} Derived from NIST SP 800-30 Table~G-5, cross-referencing the \textbf{Likelihood of Initiation} (Col 7) and \textbf{Likelihood of Success} (Col 10).
    \item \textbf{Level of Impact (Col 12):} Determined by the \textit{Unacceptable Losses} and \textit{Hazard Description}.
    \item \textbf{Risk (Col 13):} Evaluated via the NIST SP 800-30 Table~I-2 matrix by combining Overall Likelihood with Impact.
\end{itemize}

Table~\ref{tab:nist_mapping} illustrates this translation using the first row in Table~\ref{tab:hazard_h11_summary} as an example.

\begin{table*}[bthp]
  \caption{NIST SP 800-30 Table I-5 Adversarial Risk for H1.1 Attack Path 1 }
  \label{tab:nist_mapping}
  \renewcommand{\arraystretch}{1.3} % Good spacing
  \small
  \begin{tabularx}{\linewidth}{@{} p{4cm} X @{}}
    \toprule
    \textbf{NIST Parameter} & \textbf{Assessment Content} \\
    \midrule

    % 1. Event
    \textbf{1. Threat Event} & 
    The adversary first executes VM escape on Experiment Node Server (T1611), and laterally gains access to Ops Server (T1021, T1210, T1550, T1563, T1570). \\
    
    % 2. Source
    \textbf{2. Threat Source} & 
    \textit{Malicious Insider:} Authorized account with shell access on Experimental Node that they own. \\
    
    % 3. Capability
    \textbf{3. Capability} (TABLE~D-3) & 
    \textit{High:} The adversary has shell access, local network adjacency, and knowledge of the testbed architecture. 
    % (NIST SP 800-30 TABLE~D-3) 
    \\
    
    % 4. Intent
    \textbf{4. Intent} (TABLE~D-4) & 
    \textit{Malicious/Very High:} The adversary seeks to gain control of the AR-CI. \\
    
    % 5. Targeting
    \textbf{5. Targeting} (TABLE~D-5) & 
    \textit{High:} The adversary is specifically targeting the management plane (Ops Server). \\
    
    % 6. Relevance
    \textbf{6. Relevance} & 
    \textit{Confirmed:} This path leads to Hazard H1.1, and thus linked to $L1-L5$. \\
    
    % 7. Likelihood Initiation
    \textbf{7. Likelihood of Attack Initiation} (TABLE~G-2)& 
    \textit{Moderate:} Account owners can be malicious, and authorized accounts could be stolen. \\
    
    % 8. Vulns
    % \textbf{8. Vulnerabilities and Predisposing Conditions} & 
    % \textit{Vulns:} Hop1.1: \textit{T1611}: Escape to Host; \newline
    % \textit{Vulns:} Hop 1.2: \textit{{T1021}:, [T1210], [T1550], [T1563], [T1570]} 
    
    \textbf{8. Vulnerabilities and Predisposing Conditions}& 
    \textit{Vulns:} Hop1.1: \textit{T1611}: Escape to Host; \newline
    \textit{Vulns:} Hop1.2: \textit{T1021}: Remote Services, \textit{T1210}: Exploitation of Remote Services, \textit{T1550}: Use Alternate Authentication Material, \textit{T1563}: Remote Service Session Hijacking, \textit{T1570}: Lateral Tool Transfer \newline 
    \textit{Conds:} sudo privileges within Experiment Node, Experiment Node is hosted as a Virtual Machine on the Experiment Node server, out-of-band management network connecting Node Server to Ops Server.\\
    
    % 9. Severity
    \textbf{9. Severity \& Pervasiveness} (TABLE~F-2/F-3)& 
    \textit{Moderate:} The Experiment Nodes share the same software and architecture. Vulnerabilities is of moderate concern.\\
    
    % 10. Success
    \textbf{10. Likelihood of Success} & 
    \textit{Low:} Mitigated by strict protections: QEMU/KVM Virtualization and Podman containers on Experiment Node Server, SSH authentication and other security mechanisms such as patch management on Ops Server. \\
    % \textit{Low:} Mariner rootless user namespaces and QEMU/KVM virtualization enforce low conditional success probabilities at each sequential hop. \\
    
    % 11. Overall Likelihood
    \textbf{11. Overall Likelihood} (TABLE G-5) & 
    \textit{Low:} (Combination of \textit{Moderate} Initiation with \textit{Low} Success). \\
    
    % 12. Impact
    \textbf{12. Level of Impact} & 
    \textit{Very High:} Compromise of Ops Server is equivalent to breaking of Confidentiality, Integrity, and Availability, as well as suffering all unacceptable losses. \\
    
    % 13. Risk
    \textbf{13. Risk} (TABLE I-2) & 
    \textit{Moderate:} The \textit{Very High} impact is partially mediated by the \textit{Low} overall likelihood. \\
    
    \bottomrule
  \end{tabularx}
\end{table*}

\subsection{Risk-Driven Mitigation}

Due to the limited resources of AR-CI, prioritizing meaningful mitigations is essential. If the framework does not provide detailed guidelines for risk response, CISAF offers the following practical principles to help researchers navigate the risk mitigation planning:

First, \textbf{Risk Prioritization}. AR-CI teams can prioritize attack paths based on the risk of the path (combining overall likelihood and severity of the loss), and refer to the \textit{Mitigations} and \textit{Detection Strategy} section in each of the MITRE ATT\&CK~\cite{mitre_attack_2025} techniques. 

Second, \textbf{attack bottlenecks and shared vulnerabilities}. Because an adversary must complete each hop sequentially, blocking a single bottleneck breaks the entire path. Researchers can potentially develop mitigation on the common hop of multiple high risk paths and neutralize multiple risks at once. Also, blocking a single bottleneck breaks the entire attack path, so the mitigations could be implemented at a hop that convenience the team. 

Third, \textbf{mission-aware risk acceptance}. AR-CI mitigation should not break research utility. In our demonstration, vulnerabilities such as packet flooding in Experiment network may be consciously accepted rather than mitigated with intrusive controls (e.g., rate limiting or payload inspection), ensuring that security measures do not interfere with legitimate scientific experimentation.

\section{Limitations}

Our approach to CI security posture analysis has several limitations. 

\textbf{Comprehensiveness.} Although the approach aims to generate comprehensive attack graphs, many steps are either manual or generated by an LLM, and thus prone to oversight by the analysts. In the future, we suggest further automation and advancements, such as topology mapping, asset scanning and identification, vulnerability scanning, and CVE dataset integrated analysis to ensure complete graph generation and coverage.

\textbf{Protection Blind Spots.}
The presence of security mechanisms in some attack paths may create a false sense of security: having protection between two entities does not necessarily prevent the malicious party from attacking/traversing. Threat analysis frameworks, such as the STRIDE model \cite{howard2006security} or the ATT\&CK framework \cite{mitre_attack_2025} help understand the existing adversarial behaviors, but not necessarily the unknown or 0-days. 
Keeping the protective mechanisms up to date and free of known vulnerabilities is always a good practice, as security is not once and for all.
% Furthermore, the LLM is not an Oracle yet and it will make mistakes.

% For instance, Hop~1.2 from Node Server to Ops Server in Table~\ref{tab:hazard_h11_summary}, the model tagged local privilege escalation ($T_{1068}$) and host-level execution primitives ($S_{\text{hyp}}$) as \texttt{[GAP]}, despite the adversary already possessing root privileges on the Node Hypervisor and requiring lateral pivoting to Ops Server. 

% Reproducibility may be limited due to two factors: (1) the non-deterministic nature of LLM, and (2) the model may be unavailable in the future. Optimistically, we expect future LLM or other AI technology to be smarter and more accurate.
% The generated results (Appendix \ref{sec:threats}), fully reviewed by the researchers, may still contain errors or oversights. 

\textbf{Detection and Real-time Mitigation.}
Most security mechanisms, like firewalls, access control, etc., are preventative. In certain cases, preventative measures are insufficient to contain the malicious party. For example, hazard H5 can manifest as crypto-mining activities by authorized adversaries. As an insider attack happening on an Experiment Node, with no protections between the start and end nodes of the attack graph. In general, we suggest the development of detection strategies and mitigation approaches for hazards where the attack paths are short or have few or no deployable security mechanisms. 

\textbf{Risk Estimation.}
Our analysis approach produces attack graphs, which shows possible attack paths. This is necessary but not sufficient to estimate the risk of certain attacks and prioritize the deployment of new security mechanisms. 

% Future research can consider applying the Likelihood Assessment Methodology as suggested by \cite{nist_csf2_2024} and incorporating factors such as the importance of security missions, the ease of an attack, the attack's success rate and popularity, and the number of attack paths a given security mechanism can break.

\section{Conclusions}

AR-CIs are necessary to facilitate fast and equitable research progress in many scientific domains. These infrastructures face unique cybersecurity challenges that stem from their unique user population and workload, their open nature, and their small and underfunded project teams. We have proposed a simple framework to analyze the cybersecurity posture for AR-CIs, demonstrated the deliverability of porting results into a risk analysis framework, and illustrated its application to the SPHERE Testbed. We hope future research builds on this work to make it more comprehensive, data-oriented, and automated, and that it leads to improved cybersecurity of AR-CIs.

\section{LLM Usage Statement}
\label{sec:LLM}
CISAF uses a Large Language Model (LLM), specifically Google \textit{Gemini 3.7 Flash} in conjunction with \textit{Google Antigravity} \cite{google_antigravity_2}, in the process of attack path generation and the ATT\&CK technique~\cite{mitre_attack_2025} analysis:
\begin{enumerate}
\item \textbf{Attack Path Discovery:} We utilize a structured prompt,  providing information from publicly available documentation of the SPHERE testbed, the target asset(s) of the analysis, the potential starting point of the attack graphs, and the output format we want, to obtain the potential attack paths. After manually reviewing all paths, we proceed with multiple rounds of conversations with the LLM to clarify other potential paths and remove erroneous paths. Due to the stochastic nature of LLMs, generated paths may be neither exhaustive nor error-free. In practice, observed errors were primarily logically redundant and irrelevant paths rather than factually flawed (hallucinatory) content. However, based on our assumption that many AR-CI administrators are limited in security knowledge, the LLM-generated paths plus a human auditor provide fuller coverage than the result created through bare imagination and research. This task could be fully replaced if a better path generation method is present.

\item \textbf{Threats Traversal and Evaluation:} We utilize a second prompt, attached with information of the same documentation of the SPHERE testbed, the target asset(s), the formatted paths from the previous stage, and prebuilt deterministic Python tools for the LLM to extract and process the MITRE ATT\&CK STIX 2.1 dataset~\cite{mitre_attack_2025} into CSV(s).
% for the LLM to query the MITRE ATT\&CK STIX 2.1 dataset~\cite{mitre_attack_2025}, and later output results into a csv file. 

The LLM operates on an orchestrator-worker protocol, which (1) For each architectural boundary crossing (hop), an orchestrator extracts candidate ATT\&CK techniques matching the underlying operating system and device platform; (2) fires up multiple rounds of disposable clean context sub-agents to evaluate whether the ATT\&CK techniques would be \texttt{[Covered]} by the protection or a \texttt{[GAP]} for the protection until no more \texttt{[GAP]} is discovered. 

During the generation process, we notice the LLM would occasionally forget certain guidelines. When security filtering (platform halting the LLM and blocking output), shortcuts-taking (not generating sub-agents and directly spitting out results) or alignment degradation (all techniques are marked as \texttt{[GAP]}) was observed, we stopped the LLM, removed all existing outputs and started over from a fresh agent.

The LLM output is then filtered for specific per-hop ATT\&CK tactics, such as \textit{Privilege Escalation}, \textit{Lateral Movement}, \textit{Initial Access}, and \textit{Impact}. For simplicity, the 4 labels are re-categorized into 2 labels: \texttt{[Partial]} and \texttt{[GAP]} into \textit{Exposed} and \texttt{[Covered]} and \texttt{[Omitted]} into \textit{Protected}. The result is then reviewed and labeled by one of the authors. 

Across the 3 evaluated attack graphs (448 parent techniques), the LLM exhibits high fidelity when identifying defenses---5 out of 222  \textit{Protected} rows are labeled differently by the human evaluator. Conversely, the model displays huge bias for vulnerabilities---out of 226 techniques the LLM labeled as Exposed, the human evaluator labeled 124 as \textit{Protected}.
\end{enumerate}

Besides the above mentioned LLM usage, we also use LLM for editorial purposes (grammar checking and rephrasing) in this manuscript, and all outputs were inspected by the authors to ensure accuracy and originality.

% The two main challenges we found using the LLM is \textbf{(1) model consistency/stochasticity}. We tries to mitigate potential output variability through structured prompt engineering, ad-hoc human inspection and rerun, iterative multi-epoch auditing with clean context sub-agents until no new \texttt{[Partial]} or \texttt{[GAP]} found. The data is then filtered through ATT\&CK tactics on per-hop basis, as we only consider the reasonable tactics (considering privilege escalation in a lateral movement case does not make sense). We manually inspect of all post tactics-filtering threat traversal output. We uncovered that the LLM unavoidably made mistakes and inconsistencies. \textbf{(2) limited accuracy}. As mentioned in section~\ref{sec:attack_graph_llm}, the current LLM model we use, due to numerous reasons, suffer from high false positive rate (marking \textit{Protected} as \textit{Exposed}). The current stage of LLM we built provides very limited help to the researchers of the AR-CIs.

% We are able to spot and fix these inconsistencies through cross-referencing and manual analysis in our human review process. The current establishment for this project has not yet find a way to achieve high accuracy results without human proctoring and correcting during the LLM generation phase.

\bibliographystyle{IEEEtran}
\bibliography{sample-base}

% \vspace{12pt}
% \color{red}
% IEEE conference templates contain guidance text for composing and formatting conference papers. Please ensure that all template text is removed from your conference paper prior to submission to the conference. Failure to remove the template text from your paper may result in your paper not being published.

\onecolumn
\appendices

\section{Prompt for Large Language Model}

\label{sec:appendix_prompt}
\textbf{Prompt for Generate Attack Paths}
\begin{lstlisting}
# Role & Objective

You are an expert Cyberinfrastructure Security Architect conducting a systematic threat traversal analysis for a Cyberinfrastructure (CI) testbed.
Your objective is to generate an exhaustive, mathematically deduplicated set of single or multi-hop attack paths terminating at the target asset: Infrapod Server.

---

## 1. System Context & Architecture Specification

The reference implementation for a Merge testbed facility are the **Merge automated resource space (Mars)** ([mergetb/facility/mars](https://gitlab.com/mergetb/facility/mars)), topology from the graph sphere-bare_1.png, and ./Facility_Design_and_Implementation_MergeTB.html.

---

## 2. Path Derivation Protocol & Invariants

### Invariant 1: Adversary-Controlled Origins Only
Every valid attack path MUST strictly originate from a device directly accessible to an adversary:
1. `Experiment Node` (Guest VM context)
2. `Tenant Bare-Metal Node` (Root on bare-metal hardware, isolated in experiment VRF)
3. `External Internet Host` (Untrusted public network)
4. `Portal User Session / k8s container` (User container in Portal K8s cluster)
5. `Compromised Edge / Gateway Worker` (Authorized VPN/overlay termination point)
*Paths starting at internal intermediate daemons or ungrounded pods are forbidden.*

### Invariant 2: Explicit Hop Deconstruction
Deconstruct each path into discrete physical, logical, and virtualization boundary crossings:
$$\text{Origin} \xrightarrow[\text{Hop } k.1]{\text{Boundary 1}} \text{Intermediate Asset} \xrightarrow[\text{Hop } k.2]{\text{Boundary 2}} \dots \xrightarrow[\text{Hop } k.m]{\text{Boundary } m} \text{Target Asset}$$
For each hop, explicitly specify:
- Communicating Network Plane (e.g., Tenant VRF/Xpnet, Infranet, Out-of-Band Management, WireGuard/VXLAN, or Inter-Process/Hypervisor).
- Evaluation Platforms (e.g., `["Linux"]`, `["Containers", "Linux"]`, `["Network Devices", "Linux"]`).

### Invariant 3: Ockham's Razor Deduplication
Evaluate whether proposed paths share identical physical links, kernel interfaces, and daemon listening sockets. Collapse structurally identical paths into a single canonical path.

---

## 3. Required Output Format

For each discovered path, output:
1. **Path Title & Traversal Chain**: Complete end-to-end node sequence.
2. **Hop-by-Hop Breakdown**:
   - `Hop X.1`: Source $\to$ Target (Platform: `[...]`, Plane: `[...]`, Traversal Mechanism: `[...]`).
   - `Hop X.2`: Source $\to$ Target (Platform: `[...]`, Plane: `[...]`, Traversal Mechanism: `[...]`).

\end{lstlisting}
\newpage
\textbf{Prompt for Large Language Model to Generate Threat Evaluations of Listed Paths}
% \label{sec:appendix_threat_prompt}
\begin{lstlisting}
# Academic Context, Role & Defensive Scope

You are an expert Cyberinfrastructure Security Researcher conducting an empirical, evidence-grounded evaluation for a cyberinfrastructure.

Your objective is cross-referencing system configurations against the **MITRE ATT&CK Enterprise Matrix (STIX 2.1)** and the official **MergeTB Facility Architecture Specifications**, and fill in your judgement into the result csv.

---

## 1. Embedded Testbed Architecture Ground Truth (MergeTB Mars)

The reference implementation for a Merge testbed facility are the **Merge automated resource space (Mars)** ([mergetb/facility/mars](https://gitlab.com/mergetb/facility/mars)) and ./Facility_Design_and_Implementation_MergeTB.html.

### 1.1 Reconciler Architecture Pattern
- **Autonomous State Reconciliation**: The **API Server** receives an experiment materialization request from a portal, translates it into a target state specification, and writes it to **`etcd`** (and **`MinIO`** for large objects like disk images). The API server never calls facility subsystems directly.
- **Reconcilers**: Independent subsystems watch their respective key-spaces in `etcd` and drive physical/virtual resources toward the target specification.

### 1.2 The Three Facility Networks
1. **Management Network (Out-of-Band)**:
   - Dedicated physical network connecting server IPMI and management ports (`ops`, `infrapod`, `node servers`) and switch management ports.
   - Completely isolated from tenant traffic. Carries `etcd`, `MinIO`, Ground Control PXE/TFTP provisioning, and Mars reconciler channels.
2. **Infrastructure Network (Infranet)**:
   - Control-plane network dynamically orchestrated by Canopy using Linux Netlink (`rtnl`) and FRRouting (`frr`).
   - Connects facility services (`foundryd`, `cniman`, `mariner`) to gateway workers and node hypervisors.
3. **Experiment Network (Xpnet / VRF)**:
   - Isolated data-plane network instantiated per experiment materialization.
   - Enforced via VXLAN overlays and VLAN-aware bridges. Carries untrusted tenant traffic and connects Experimental Nodes directly to their dedicated Infrapod Pod (`dance`/`foundryd`).

### 1.3 Core Mars Services & Microservices
1. **`apiserver`** ([service/apiserver](https://gitlab.com/mergetb/facility/mars/-/tree/main/service/apiserver)):
   - Implements MergeTB Facility API (`Materialize`, `Dematerialize`). Writes target states to `etcd` / `MinIO`.
2. **`infrapod`** ([service/infrapod](https://gitlab.com/mergetb/facility/mars/-/tree/main/service/infrapod)):
   - Per-materialization **Podman pods** created via **Cniman** for custom CNI networking.
   - Runs 4 microservice containers:
     - **`dance`** ([service/dance](https://gitlab.com/mergetb/facility/mars/-/tree/main/service/dance)): In-memory DHCP and DNS server watching `etcd`.
     - **`foundry`** ([tech/foundry](https://gitlab.com/mergetb/tech/foundry)): Node configuration service. The server (`foundryd`) listens on the experiment VRF (only reachable by nodes in that experiment). When the node client (`foundryc`) boots, it connects to `foundry` (resolved by `dance`) to receive SSH keys, accounts, and network configuration.
     - **`moactl`**: Network emulation controller.
     - **`podinit`** ([service/podinit](https://gitlab.com/mergetb/facility/mars/-/tree/main/service/podinit)): Initial bootstrapping container configuring routing and NAT inside the pod network namespace.
3. **`canopy`** ([service/canopy](https://gitlab.com/mergetb/facility/mars/-/tree/main/service/canopy)):
   - Virtual network reconciler running on switches (Cumulus Linux, DENT, Switch Linux), infrastructure servers, network emulators, and hypervisors.
   - Configures VLAN access/trunk ports, VLAN subinterfaces, VLAN-aware bridging, VXLAN VTEPs, and BGP Router specifications (EVPN) using the **`rtnl`** Netlink socket library ([tech/rtnl](https://gitlab.com/mergetb/tech/rtnl)).
4. **`frr`**: Routing engine used by Canopy via `vtysh` CLI / Unix domain sockets.
5. **`mariner`** ([service/mariner](https://gitlab.com/mergetb/facility/mars/-/tree/main/service/mariner)):
   - Node virtualization reconciler provisioning **Qemu/KVM virtual machines** inside standalone Podman containers (q35 chipset; `virtio`, `e1000`, `e1000e` TAP interfaces connected to Canopy bridges).
6. **`sled`** ([service/sled](https://gitlab.com/mergetb/facility/mars/-/tree/main/service/sled)):
   - Bare-metal OS imaging system. Client is a **u-root** microkernel delivered via DHCP/PXE/TFTP/HTTP, streaming disk images from MinIO and executing `kexec`.
7. **`metal`** ([service/metal](https://gitlab.com/mergetb/facility/mars/-/tree/main/service/metal)): Power control (APC PDUs, IPMI, NCD Relay Boards, Raven virtual power).
8. **`tftp`** ([service/tftp](https://gitlab.com/mergetb/facility/mars/-/tree/main/service/tftp)): Static boot images on port 69 on the shared service VRF.
9. **`Ground Control` / `ops`**: Zero-Touch Provisioning (DHCP, DNS, PXE, Ignition) on the management network for **Fedora CoreOS (FCOS)** infrastructure servers and **Cumulus Linux** switches.

---

## 2. Testbed Topology & Physical/Logical Interconnects

```mermaid
flowchart TD
    subgraph External_Zone["External Zone"]
        Internet["External Internet / Border"]
        User["AR-CI User / Researcher Workstation"]
    end

    subgraph Portal_Zone["Portal Infrastructure (K8s Cluster)"]
        Bastion["Bastion Server (SSH Gateway)"]
        Gateway["Portal Gateway Worker (K8s Pod / Wireguard)"]
        PortalAPI["Portal API & Identity Server"]
    end

    subgraph Facility_Switches["Facility Switching Fabric"]
        IngressSW["Ingress / Border Switch"]
        MgmtSW["Management Switch (Out-of-Band Ops Network)"]
        InfranetSW["Infranet Leaf Switch (Control Network)"]
        XpnetSW["Xpnet Leaf Switch (Tenant Experiment VRF / VXLAN)"]
    end

    subgraph Facility_Servers["Facility Compute & Infrastructure Servers"]
        Ops["Operations Server (ops)\n- Ground Control (DHCP/DNS/PXE)\n- Mars Reconciler & etcd/MinIO"]
        
        subgraph Infrapod_Host["Infrapod Host Server (Fedora CoreOS)"]
            PodmanHost["Host OS (systemd / Cniman / rootless Podman)"]
            subgraph Infrapod_Pod["Infrapod Pod (Per-Materialization)"]
                Dance["dance (DHCP/DNS)"]
                Foundry["foundryd (Node Config)"]
                Moa["moactl (Emulation)"]
                Podinit["podinit (Routing/NAT)"]
            end
        end

        subgraph Node_Server_1["Node Server 1 (Bare-Metal / Mariner Hypervisor)"]
            Host1["Host Hypervisor (Fedora CoreOS / Sled)"]
            ExpNode1["Experimental Node 1 (VM or Bare-Metal Tenant Context)"]
        end
    end

    %% Interconnects
    User -->|SSH / HTTPS| Bastion
    Bastion --> Gateway
    Internet -->|Border Ingress| IngressSW
    IngressSW --> Infrapod_Host
    
    Gateway -->|VXLAN / Infranet| InfranetSW
    InfranetSW --> Infrapod_Host
    InfranetSW --> Node_Server_1
    
    ExpNode1 <-->|Tenant VRF / TAP| XpnetSW
    XpnetSW <-->|Canopy VXLAN Bridge| Infrapod_Pod
    
    MgmtSW -.->|Out-of-Band Ops| Ops
    MgmtSW -.->|Out-of-Band Ops| Infrapod_Host
    MgmtSW -.->|Out-of-Band Ops| Node_Server_1
```

---

## 3. Table 3 Threat Traversal Paths (Hazard H1.1 Boundary Assessment)

Hazard H1.1 evaluates the architectural boundary integrity of the Infrapod subsystem to verify isolation of tenant experiment traffic:

# ==============================================================================
# MARS / MERGETB CYBERINFRASTRUCTURE TESTBED: ATTACK PATHS TO OPS SERVER
# Target Asset: Ops Server (Facility Core: API Server, etcd, MinIO, Sled, Metal)
# Deduplication: Ockham's Razor applied across kernel interfaces & listening sockets
# ==============================================================================

### Path 1: VM Hypervisor Escape & Infranet Management Plane Pivot
Traversal Chain:
Experiment Node -> Node Server (Hypervisor) -> Ops Server

Hop-by-Hop Breakdown:
- Hop 1.1: Experiment Node -> Node Server (Hypervisor)

- Hop 1.2: Node Server (Hypervisor) -> Ops Server
  * Platform: ["Linux", "Network Devices"]

--------------------------------------------------------------------------------

### Path 2: Bare-Metal Switch VLAN Hopping & Shared Service Plane Pivot
Traversal Chain:
Tenant Bare-Metal Node -> Facility Access Switch -> Ops Server

Hop-by-Hop Breakdown:
- Hop 2.1: Tenant Bare-Metal Node -> Facility Access Switch
  * Platform: ["Network Devices", "Linux"]

- Hop 2.2: Facility Access Switch -> Ops Server
  * Platform: ["Network Devices", "Linux"]

--------------------------------------------------------------------------------

### Path 3: Per-Materialization Infrapod Subversion & Multi-Hop Facility Pivot
Traversal Chain:
Tenant Node (VM or Bare-Metal) -> Infrapod Container -> Infrapod Host Server -> Ops Server

Hop-by-Hop Breakdown:
- Hop 3.1: Tenant Node -> Infrapod Container (dance / foundry)
  * Platform: ["Containers", "Linux"]

- Hop 3.2: Infrapod Container (dance / foundry) -> Infrapod Host Server
  * Platform: ["Containers", "Linux"]

- Hop 3.3: Infrapod Host Server -> Ops Server
  * Platform: ["Linux", "Network Devices"]

--------------------------------------------------------------------------------

### Path 4: External Boundary Ingress & Mars API Server Access
Traversal Chain:
External Internet Host -> Facility Ingress Switch / Border Gateway -> Ops Server

Hop-by-Hop Breakdown:
- Hop 4.1: External Internet Host -> Facility Ingress Switch / Border Gateway
  * Platform: ["Network Devices", "Linux"]

- Hop 4.2: Facility Ingress Switch / Border Gateway -> Ops Server
  * Platform: ["Network Devices", "Linux"]


--------------------------------------------------------------------------------

### Path 5: Portal k8s Container / Gateway Worker Cross-Facility Control Plane Pivot
Traversal Chain:
Portal User Session / k8s Container -> Portal Gateway Worker -> Ops Server

Hop-by-Hop Breakdown:
- Hop 5.1: Portal User Session / k8s Container -> Portal Gateway Worker
  * Platform: ["Containers", "Linux"]

- Hop 5.2: Portal Gateway Worker -> Ops Server
  * Platform: ["Linux", "Network Devices"]


---

## 4. The 9 Operational Guardrails & Invariants

- **Guardrail 1: 3-Layer Platform Deconstruction**: Decompose every target component across Physical/Host, Encapsulation/Runtime, and OS/Userland based on the MITRE ATT&CK provide `x_mitre_platforms`. (e.g. Infrapod Pod = `["Containers", "Linux"]`; Infranet Switch = `["Network Devices", "Linux"]`).
- **Guardrail 2: Data-First STIX Platform Discovery (Zero Phantom Tactics)**: Query STIX for tactics present on target platforms. For tactics that exist on the platform but are non-applicable for a hop, emit an explicit summary row (`Txxxx` or `Txxxx.yyy`, `[Tactic Name] (Tactic Omitted)`, `[Omitted / Out of Scope]`). 
- **Guardrail 3: Programmatic Full-Table Audit**: Process 100% of candidate techniques for the target platforms. Manual spot-checking is prohibited.
- **Guardrail 4: Sub-technique Auto-Collapsing**: Collapse fully covered parent+subtechniques into parent rows (annotating collapsed count). If any sub-technique is `[GAP]` or `[Partial]`, expand all sub-techniques of the corresponding parent technique. Executed via `scripts/collapse_subtechniques.py`.
- **Guardrail 5: Verifiable Documentation Quotes & Citations**: Every row must include direct verbatim quotes from `mergetb_facility_architecture_docs.md` or any other trusted sources approved by the user, with section citations in `Testbed Evidence & Documentation Quotes`. Searching of sources is allowed when information provided is not enough, but explicitly requires user review and confirmation for the new source to be applicable for inclusion and analysis.
- **Guardrail 6: Simple In-Place Status Schema**: Maintain a single clean status column: **`SPHERE Protection Status`** (`[Covered]`, `[Partial]`, `[GAP]`, or `[Omitted / Out of Scope]`). Update in place upon epoch transitions.
- **Guardrail 7: Strict STIX Mitigation Decoupling**: `MITRE Mitigations` contains strictly native STIX `course-of-action` (`Mxxxx`) objects. Network boundary isolation belongs in `SPHERE Mechanism & Coverage Rationale`.
- **Guardrail 8: Strict Anti-Hardcoding & Zero-Rule-Script Invariant**:
  - Python scripts are strictly restricted to **Generic Dumb I/O Tools** (`scripts/export_stix_skeleton.py`, `scripts/update_audit_table.py`, `scripts/compute_delta.py`, `scripts/collapse_subtechniques.py`).
  - **BANNED**: Writing any script (e.g., `audit_*.py`) that contains hardcoded ATT&CK IDs (`Txxxx`), heuristic `if/elif` threat rules, static status assignments, or pre-baked rationale strings.
  - **MANDATORY**: 100% of qualitative threat evaluations, boundary traversal analyses, and documentation quote selections MUST be performed dynamically by the LLM Agent in-context.
- **Guardrail 9: Stateless Multi-Epoch Forking (Zero Context Pollution)**: Do not iterate multi-pass epochs within a dirty context. For each epoch $i$, the parent process forks a clean-context child subagent (`CHILD_EPOCH_WORKER`) passing explicitly $\mathcal{G}_{i-1}$. The child evaluates the epoch, writes `data/Table4_PathX_HopY_ATTACK_Audit(Epoch_i).csv`, outputs $\mathcal{G}_i$, and terminates immediately (no reuse). The parent computes $\Delta_i = \mathcal{G}_i \setminus \mathcal{G}_{i-1}$ until $\Delta_i = \emptyset$.

---

## 5. Generic Data Tools & Dumb I/O Scripts

All scripts are completely domain-agnostic and contain 0% hardcoded security evaluation rules:

1. **Universal STIX Skeleton Exporter**:
   ```bash
   python3 scripts/export_stix_skeleton.py --platforms Containers Linux --path-id "Path 1" --hop "Hop 1.1: Experiment Node -> Infrapod Pod" --output data/Path1_Hop1.1_Skeleton.csv
   ```
   *(Extracts all candidate techniques and native STIX `Mxxxx` mitigations with 0% hardcoding).*

2. **STIX Query Tool**:
   ```bash
   python3 scripts/query_attack.py --platform Containers --tactic execution
   ```

3. **Generic Audit Table Updater (Dumb I/O)**:
   ```bash
   python3 scripts/update_audit_table.py --input-csv data/Path1_Hop1.1_Skeleton.csv --updates-json scratch/chunk_eval.json --output-csv data/Table4_Path1_Hop1.1_ATTACK_Audit(Epoch_1).csv
   ```
   *(Takes an LLM-evaluated JSON payload and applies it to the target CSV).*

4. **Epoch Gap Set & Delta Convergence Computer**:
   ```bash
   python3 scripts/compute_delta.py --current-epoch-csv data/Table4_Path1_Hop1.1_ATTACK_Audit(Epoch_1).csv
   python3 scripts/compute_delta.py --current-epoch-csv data/Table4_Path1_Hop1.1_ATTACK_Audit(Epoch_2).csv --prev-epoch-csv data/Table4_Path1_Hop1.1_ATTACK_Audit(Epoch_1).csv
   ```
   *(Calculates $\mathcal{G}_i$ and $\Delta_i = \mathcal{G}_i \setminus \mathcal{G}_{i-1}$ using pure set math on CSV status columns).*

5. **Sub-Technique Collapsing Tool (Guardrail 4)**:
   ```bash
   python3 scripts/collapse_subtechniques.py --input-csv data/Table4_Path1_Hop1.1_ATTACK_Audit(Epoch_2).csv --output-csv data/Table4_Path1_Hop1.1_ATTACK_Audit_Final.csv
   ```

6. **Epoch reanalysis rules (Guardrail 5)**:
   When analyzing in a new epoch, you should take current [GAP], and [Partial] into consideration, and tell if they opens up new opportunities of gaining access, which potentially change tactics from [Partical] to [GAP], [Covered] to [Partial], or [Covered] to [GAP].
   
---

## 6. Forked Parent-Child Subagent Execution Protocol

To maintain 100% clean context across multi-pass epochs, avoid attention drift, and prevent confirmation bias, execution is split into a **Parent Orchestrator** and **Disposable Child Epoch Workers**:

### 6.1 Process Role & Fork Header
Every agent invocation begins with an explicit role identifier header:
```yaml
# FORK_HEADER
PROCESS_ROLE: [PARENT_ORCHESTRATOR | CHILD_EPOCH_WORKER]
TARGET_PATH: "Path 1"
TARGET_HOP: "Hop 1.1: Experiment Node -> Infrapod Pod"
CURRENT_EPOCH: 1
INPUT_GAP_SET (G_{i-1}): [] # Passed explicitly by parent to child
```

### 6.2 State Lifecycle & Division of Responsibility
```mermaid
flowchart TD
    subgraph Parent_Process["Parent Orchestrator Process"]
        P1["1. Export STIX Skeleton\n(export_stix_skeleton.py)"]
        P2["2. Spawn Child Subagent for Epoch 1\n(invoke_subagent)"]
        P3["3. Compute Delta_1 = G_1 \\ 0\n(compute_delta.py)"]
        P4{"Is Delta_i == 0?"}
        P5["4. Spawn Disposable Child for Epoch i+1\n(Passing G_i in prompt)"]
        P6["5. Compute Delta_{i+1}\n(compute_delta.py)"]
        P7["6. Apply Guardrail 4 Collapsing & Finalize CSV\n(collapse_subtechniques.py)"]
    end

    subgraph Child_Process_1["Disposable Child Worker (Epoch 1)"]
        C1_1["Clean Context Init"] --> C1_2["In-Context Baseline Audit\n(100% Candidate Techniques)"]
        C1_2 --> C1_3["Export Epoch 1 CSV & Output G_1"]
        C1_3 --> C1_4["TERMINATE IMMEDIATELY (Exit)"]
    end

    subgraph Child_Process_2["Disposable Child Worker (Epoch 2)"]
        C2_1["Clean Context Init with G_1"] --> C2_2["Ingest G_1 as Boundary Preconditions"]
        C2_2 --> C2_3["Audit Cascaded Technique Bypasses"]
        C2_3 --> C2_4["Export Epoch 2 CSV & Output G_2"]
        C2_4 --> C2_5["TERMINATE IMMEDIATELY (Exit)"]
    end

    P2 -.-> Child_Process_1
    Child_Process_1 -.->|Returns G_1| P3
    P3 --> P4
    P4 -- "No (Delta_1 > 0)" --> P5
    P5 -.-> Child_Process_2
    Child_Process_2 -.->|Returns G_2| P6
    P6 --> P4
    P4 -- "Yes (Delta_i == 0)" --> P7
```

### 6.3 Parent Orchestrator Protocol
1. **Export Skeleton**: Run `python3 scripts/export_stix_skeleton.py` for the target hop.
2. **Spawn Child Worker (Epoch 1)**: Invoke a subagent with `PROCESS_ROLE: CHILD_EPOCH_WORKER`, `CURRENT_EPOCH: 1`, `INPUT_GAP_SET: []`.
3. **Evaluate $\Delta_i$**: Upon child termination, run `scripts/compute_delta.py` to extract $\mathcal{G}_i$ and compute $\Delta_i = \mathcal{G}_i \setminus \mathcal{G}_{i-1}$.
4. **Loop or Finalize**:
   - If $\Delta_i \neq \emptyset$: Spawn a NEW, disposable child subagent for Epoch $i+1$ with clean context, passing `INPUT_GAP_SET: G_i`.
   - If $\Delta_i = \emptyset$: Run `scripts/collapse_subtechniques.py` to produce `data/Table4_PathX_HopY_ATTACK_Audit_Final.csv`.

### 6.4 Child Epoch Worker Protocol
1. **Zero-Contamination Start**: Operates in a fresh context window with no prior conversational residue.
2. **Ingest Preconditions**: Ingests $\mathcal{G}_{i-1}$ from the prompt header as pre-existing unmitigated baseline capabilities.
3. **In-Context Qualitative Evaluation**: Evaluates 100% candidate techniques against `mergetb_facility_architecture_docs.md`, determining `[Covered]`, `[Partial]`, `[GAP]`, or `[Omitted / Out of Scope]`, documenting verbatim citations and prerequisites (`[TAxxxx: Txxxx]`).
4. **Output & Immediate Self-Termination**: Writes `data/Table4_PathX_HopY_ATTACK_Audit(Epoch_i).csv`, sends a structured summary of $\mathcal{G}_i$ to Parent, and exits immediately. Child agents are NEVER reused.

---

## 7. Standard 13-Column CSV Schema
```
1. Path ID
2. Hop
3. ATT&CK ID
4. Technique Name
5. Is Subtechnique
6. Platforms
7. Evaluated Tactic(s)
8. SPHERE Protection Status  ([Covered] | [Partial] | [GAP] | [Omitted / Out of Scope])
9. MITRE Mitigations
10. SPHERE Mechanism & Coverage Rationale  (Preconditions, boundary analysis, cascade reasoning)
11. Testbed Evidence & Documentation Quotes  (Verbatim quotes & section citations)
12. ATT&CK URL
13. Prerequisite
```

---

## 8. Master Prompt Templates

### Parent Prompt Template
```markdown
# FORK_HEADER
PROCESS_ROLE: PARENT_ORCHESTRATOR
TARGET_PATH: "<Path ID>"
TARGET_HOP: "<Hop Description>"
CURRENT_EPOCH: 0
INPUT_GAP_SET: []

You are the Parent Orchestrator. Execute Step 1 (export skeleton) and manage disposable Child Epoch Workers until Delta_i == 0.
```

### Child Prompt Template (Epoch i)
```markdown
# FORK_HEADER
PROCESS_ROLE: CHILD_EPOCH_WORKER
TARGET_PATH: "<Path ID>"
TARGET_HOP: "<Hop Description>"
CURRENT_EPOCH: <i>
INPUT_GAP_SET: <G_{i-1} Array>

You are a disposable Child Epoch Worker. Perform 100% in-context audit for Epoch <i> against mergetb_facility_architecture_docs.md. Ingest INPUT_GAP_SET as pre-existing boundary conditions. Export CSV to data/Table4_<Path>_<Hop>_ATTACK_Audit(Epoch_<i>).csv, report G_<i>, and terminate immediately.
```



\end{lstlisting}

\end{document}